\documentclass[aps,pra,10pt, twocolumn, groupedaddress, superscriptaddress, amsmath, amssymb]{revtex4-1}
\usepackage[british]{babel}
\usepackage{graphicx}   
\usepackage{physics} 
\usepackage{dcolumn}
\usepackage{bm}         
\usepackage{verbatim}  
\usepackage{color}
\usepackage[dvipsnames]{xcolor}
\usepackage{diagbox} 

\usepackage{xcolor}	 
\definecolor{myblue}{RGB}{34,31,150}

\usepackage[breaklinks=true,colorlinks=true,linkcolor=myblue,urlcolor=myblue,citecolor=myblue]{hyperref}

\makeatletter
\newcommand*{\transpose}{%
  {\mathpalette\@transpose{}}%
}
\newcommand*{\@transpose}[2]{%
  \raisebox{\depth}{$\m@th#1\intercal$}%
}
\makeatother

\begin{document}

% Title, authors, research centres and date

\title{Bayesian multi-parameter quantum metrology with limited data}
\author{Jes\'{u}s Rubio}
\email{J.Rubio-Jimenez@exeter.ac.uk}
\affiliation{Department of Physics and Astronomy, University of Sussex, Brighton BN1 9QH, UK}
\affiliation{Department of Physics and Astronomy, University of Exeter, Stocker Road, Exeter EX4 4QL, UK}
\author{Jacob Dunningham}
\email{J.Dunningham@sussex.ac.uk}
\affiliation{Department of Physics and Astronomy, University of Sussex, Brighton BN1 9QH, UK}
\date{\today}
   
% Abstract

\begin{abstract}

A longstanding problem in quantum metrology is how to extract as much information as possible in realistic scenarios with not only multiple unknown parameters, but also limited measurement data and some degree of prior information. Here we present a practical solution to this: we derive a new Bayesian multi-parameter quantum bound, construct the optimal measurement when our bound can be saturated for a single shot and consider experiments involving a repeated sequence of these measurements. Our method properly accounts for the number of measurements and the degree of prior information, and we illustrate our ideas with a qubit sensing network and a model for phase imaging, clarifying the non-asymptotic role of local and global schemes. Crucially, our technique is a powerful way  of implementing quantum protocols in a wide range of practical scenarios that tools such as the Helstrom and Holevo Cram\'{e}r-Rao bounds cannot normally access.

\end{abstract}

\maketitle

% Body of the document

%-------------------------------------------------------------------------

\section{Introduction}

Real-world applications typically give rise to estimation problems with several unknown pieces of information. For instance, we may wish to determine the range and velocity of a moving object \cite{zhuang2017}, quantify phases and phase diffusion \cite{vidrighin2014, szczykulska2017}, reconstruct an image \cite{humphreys2013, knott2016local, zhang_lu2017}, estimate the components of some field \cite{baumgratz2016}, assess spatial deformations in a grid of sources \cite{jasminder2016, jasminder2018} or use quantum networks to implement distributed sensing protocols \cite{proctor2017networked, proctor2017networkedshort, ge2018, eldredge2018, altenburg2018, qian2019, guo2019}. In this context, many results in existing literature rely on the multi-parameter Cram\'{e}r-Rao bound and its quantum extension by Helstrom \cite{helstrom1976, paris2009, Szczykulska2016, sammy2016compatibility, pezze2017simultaneous, liu2019}.

This framework is powerful because sometimes there is a quantum strategy for which the classical and quantum versions of the bound coincide \cite{sammy2016compatibility, pezze2017simultaneous} (e.g., for commuting generators and pure states \cite{sammy2016compatibility, pezze2017simultaneous, proctor2017networked}). However, to exploit it one first needs to reach the classical bound. One possibility is assuming locally unbiased estimators \cite{fraser1964}, which may be reasonable for large prior information \cite{hall2012, rafal2015, haase2018jul}, although in general many repetitions of the experiment are required to approach the bound \cite{kay1993}. While this may generate fundamental results locally or at least asymptotically, the fact that the measurement data can be limited in practice (e.g., \cite{berchera2019, polino2018}) and that our prior knowledge may be moderate motivates the search for a more generally applicable strategy \cite{jesus2018, jesus2019thesis}.

Current research is exploring the Holevo Cram\'{e}r-Rao bound \cite{holevo2011, sammy2016compatibility, yang2019, carollo2019, tsang2019dec, albarelli2019novA, vidarte2019, albarelli2019novB, albarelli2019, sidhu2019dec}, which is more informative than Helstrom's counterpart (albeit moderately \cite{carollo2019, albarelli2019novA, tsang2019dec}) when the latter produces incompatible estimators. Unfortunately, this bound is still restricted by either local unbiasedness or, generally, the requirement of an asymptotically large number of copies of the scheme \cite{holevo2011, yang2019, sammy2016compatibility, tsang2019dec}. Hence, it lacks the type of generality that we seek.

Instead, we recall that the fundamental equations that the optimal quantum strategy must satisfy for a single or several copies have been known since the Bayesian works of Helstrom, Holevo, Personick, Yuen and others \cite{personick1971, helstrom1976, helstrom1974, holevo1973b, holevo1973, yuen1973}, and that the Bayesian framework does provide the tools for scenarios with limited data where the prior information plays an active role \cite{jaynes2003, jesus2019thesis}. A recent review of this formalism, including both classical and quantum aspects, can be found in chapter 3 of \cite{jesus2019thesis}.

Except for a few cases such as those admitting covariant measurements \cite{ariano1998, chiara2003, chiribella2005, holevo2011, rafal2015}, solving these equations exactly is challenging \cite{helstrom1976}, and the known solutions usually assume no a priori knowledge, with exceptions such as the single-parameter work \cite{demkowicz2011}. Fortunately, our single-parameter proposal in \cite{jesus2018} already shows that this formalism can be exploited in a less general but more practical way. In particular, if using the square error is justified (as it is for moderate prior knowledge \cite{rafal2015, jesus2017, friis2017, jesus2018, jesus2019thesis}), then we may calculate the single-shot optimal quantum strategy \cite{personick1971, macieszczak2014bayesian} and repeat it as many times as the application at hand demands or allows for \cite{jesus2018}. This generates uncertainties that have been optimised in a shot-by-shot fashion, and that sometimes recover the single-parameter quantum Cram\'{e}r-Rao bound asymptotically as a limiting case \cite{jesus2019thesis}.

The aim of this paper is to construct a multi-parameter version of the efficient shot-by-shot technique described above, a step that will generalise and take quantum metrology to  a new level by providing the means of addressing practical problems beyond the scope of the Helstrom and Holevo Cram\'{e}r-Rao bounds. First we derive a new multi-parameter quantum bound for the quadratic error in Sec. \ref{newbound}. This bound can incorporate prior information explicitly without imposing a particular form for the prior probability, and its derivation does not involve unbiasedness conditions. We then study its potential saturability for a single shot, and we discuss how and under which circumstances we can adapt this result to exploit it in strategies where the same experiment is repeated several times. To illustrate our ideas, in Sec. \ref{app} we apply our new multi-parameter technique to a qubit sensing network and a discrete model for phase imaging, and we analyse the role of \emph{local} and \emph{global} sensing protocols (in the sense of \cite{proctor2017networked, proctor2017networkedshort}) when the scheme operates in the non-asymptotic regime of a finite and possibly small number of experiments. Finally, the merits and potential extensions of our results are discussed in Sec. \ref{conclusions}.

\noindent{\it Note added.}  After completion of the first version of this work (see \footnote{The first version of the present work appeared in arXiv:1906.04123.}), Sidhu and Kok \cite{jasminder2019} arrived at the quantum bound in Sec. \ref{derivation} independently by using a geometric approach. In addition, a version of the same result and a simplification for Gaussian priors was later discussed by Demkowicz-Dobrza\ifmmode \acute{n}\else \'{n}\fi{}ski \emph{et al.} in their review \cite{rafal2020}. The interested readers will find in these works other perspectives that complement our findings, which here are instead understood in terms of the non-asymptotic version of quantum metrology developed in \cite{jesus2019thesis}. 

%-------------------------------------------------------------------------

\section{A new multi-parameter technique}\label{newbound}

\subsection{Derivation of the bound}\label{derivation}

Suppose we encode the unknown parameters $\boldsymbol{\theta} = (\theta_1, \cdots, \theta_d)$ in the probe state $\rho_0$, so that the transformed state is $\rho(\boldsymbol{\theta})$, and that we perform a single measurement $E(m)$ with outcome $m$. Then the likelihood is $p(m|\boldsymbol{\theta}) = \mathrm{Tr}[E(m) \rho(\boldsymbol{\theta})]$, and by combining it with the prior $p(\boldsymbol{\theta})$ into the joint probability $p(\boldsymbol{\theta}, m) = p(\boldsymbol{\theta}) p(m|\boldsymbol{\theta})$ we can construct the uncertainty
\begin{equation}
\bar{\epsilon}_{\mathrm{mse}} = \sum_{i=1}^d w_i \int d\boldsymbol{\theta} dm ~p(\boldsymbol{\theta}, m) \left[g_i(m) - \theta_i  \right]^2,
\label{msegen}
\end{equation}
where $g_i(m)$ is the estimator for the $i$-th parameter, $w_i \geqslant 0$ indicates its relative importance \cite{proctor2017networked} and $\sum_{i=1}^d w_i = 1$.

Let us rewrite Eq.\hspace{0.2em}(\ref{msegen}) as $\bar{\epsilon}_{\mathrm{mse}} = \mathrm{Tr}[\mathcal{W} \Sigma_{\mathrm{mse}}]$, where $\mathcal{W} = \mathrm{diag}(w_1, \dots, w_d)$ and
\begin{equation}
\Sigma_{\mathrm{mse}} =  \int d\boldsymbol{\theta} dm ~p(\boldsymbol{\theta}, m) \left[\boldsymbol{g}(m) - \boldsymbol{\theta} \right] \left[\boldsymbol{g}(m) - \boldsymbol{\theta} \right]^\transpose,
\end{equation}
and with $\boldsymbol{g}(m) = (g_1(m), \dots, g_d(m))$. The first step is performing a classical optimisation over all possible estimators. We start by constructing the scalar
\begin{equation}
\boldsymbol{u}^\transpose \Sigma_{\mathrm{mse}} \boldsymbol{u} = \int d\boldsymbol{\theta} dm ~p(\boldsymbol{\theta}, m) \left[g_u(m) - \theta_u  \right]^2,
\label{scalarquantity}
\end{equation}
with $g_u(m) = \boldsymbol{u}^\transpose \boldsymbol{g}(m) = \boldsymbol{g}^\transpose(m)\hspace{0.15em} \boldsymbol{u}$, $\theta_u = \boldsymbol{u}^\transpose \boldsymbol{\theta} = \boldsymbol{\theta}^\transpose \boldsymbol{u}$ and $\boldsymbol{u}$ being an arbitrary real vector. If we look at $\boldsymbol{u}^\transpose \Sigma_{\mathrm{mse}} \boldsymbol{u}$ as a functional of $g_u(m)$ \cite{jaynes2003, jesus2017}, that is, $\boldsymbol{u}^\transpose \Sigma_{\mathrm{mse}} \boldsymbol{u} = \epsilon\left[g_u(m)\right]$, then we can formulate the variational problem
\begin{equation}
\delta \epsilon\left[g_u(m)\right] = \delta \int dm~\mathcal{L}\left[m, g_u(m) \right] = 0,
\label{variationalprob}
\end{equation} 
with $\mathcal{L}\left[m, g_u(m) \right] = \int d\boldsymbol{\theta} p(\boldsymbol{\theta}, m) \left[g_u(m) - \theta_u  \right]^2 $. Its solution, which we revisit in Appendix \ref{classicalest}, is $g_u(m) = \int d\boldsymbol{\theta} p(\boldsymbol{\theta}|m)\theta_u$, where $p(\boldsymbol{\theta}|m) \propto p(\boldsymbol{\theta})p(m|\boldsymbol{\theta})$ is the posterior probability. 

The previous calculation implies that the vector estimator that makes the uncertainty extremal is $\boldsymbol{g}(m) =  \int d\boldsymbol{\theta} p(\boldsymbol{\theta}|m)\boldsymbol{\theta}$, and this is precisely the solution that is known to achieve the minimum matrix error (see \cite{kay1993} and Appendix \ref{classicalest}). Hence, we have that $\boldsymbol{u}^\transpose \Sigma_{\mathrm{mse}} \boldsymbol{u} \geqslant \boldsymbol{u}^\transpose \Sigma_c \boldsymbol{u}$ after introducing $\boldsymbol{g}(m) =  \int d\boldsymbol{\theta} p(\boldsymbol{\theta}|m)\boldsymbol{\theta}$ in Eq.\hspace{0.2em}(\ref{scalarquantity}), where
\begin{eqnarray}
\boldsymbol{u}^\transpose \Sigma_c \boldsymbol{u} = \hspace{-0.4em}\int \hspace{-0.15em} d\boldsymbol{\theta} p(\boldsymbol{\theta})\theta_u^2 
-\hspace{-0.4em} \int \hspace{-0.15em} dm  \frac{\left[ \int d\boldsymbol{\theta} p(\boldsymbol{\theta})p(m|\boldsymbol{\theta}) \theta_u \right]^2}{\int d\boldsymbol{\theta} p(\boldsymbol{\theta})p(m|\boldsymbol{\theta})}.
\label{optclassical}
\end{eqnarray}

Next we examine the quantum part of the problem. By inserting $p(m|\boldsymbol{\theta}) = \mathrm{Tr}[E(m) \rho(\boldsymbol{\theta})]$ in Eq.\hspace{0.2em}(\ref{optclassical}),
\begin{equation}
\boldsymbol{u}^\transpose \Sigma_c \boldsymbol{u} = \int d\boldsymbol{\theta} p(\boldsymbol{\theta})\theta_u^2 - \int dm  \frac{\mathrm{Tr}\left[ E(m) \bar{\rho}_u \right]^2}{\mathrm{Tr}\left[ E(m) \rho \right]},
\label{bayesanalogy}
\end{equation}
where $\rho = \int d\boldsymbol{\theta} p(\boldsymbol{\theta}) \rho(\boldsymbol{\theta})$ and $\bar{\rho}_u = \int d\boldsymbol{\theta} p(\boldsymbol{\theta}) \rho(\boldsymbol{\theta}) \theta_u$. Remarkably, the second term is formally analogous to the result of applying the Born rule to the expression for the classical Fisher information \footnote{More concretely, $F(\theta) = \int dm [\partial_\theta p(m|\theta)]^2/p(m|\theta) = \int dm \mathrm{Tr}[E(m)\partial_\theta\rho(\theta)]^2/\mathrm{Tr}[E(m)\rho(\theta)]$ \cite{BraunsteinCaves1994, genoni2008}.}. This suggests the possibility of bounding it with a procedure similar to the proof of the Braunstein-Caves inequality \cite{BraunsteinCaves1994, genoni2008}.  

Following this analogy we introduce the Bayesian counterpart of the equation for the symmetric logarithmic derivative \footnote{The symmetric logarithmic derivative $L(\theta)$, which is a notion associated with the quantum Fisher information, is given by $L(\theta)\rho(\theta) + \rho(\theta)L(\theta) = 2 \partial_\theta \rho(\theta)$ \cite{rafal2015, paris2009}.}, i.e., $S_u \rho + \rho S_u = 2\bar{\rho}_u$. This allows us to manipulate the second term in Eq.\hspace{0.2em}(\ref{bayesanalogy}), which we denote by $B_u$, as 
\begin{align}
B_u &= \int dm\left(\frac{\mathrm{Re}\left\lbrace\mathrm{Tr}\left[E(m) S_u\rho\right]\right\rbrace}{\sqrt{\mathrm{Tr}\left[ E(m) \rho\right]}}\right)^2 
\nonumber \\
&\leqslant \int dm~\abs{\frac{\mathrm{Tr}\left[E(m) S_u\rho\right]}{\sqrt{\mathrm{Tr}\left[ E(m) \rho\right]}}}^2
\nonumber \\
&=\int dm~ \abs{\mathrm{Tr}\left[\frac{\rho^{\frac{1}{2}}E(m)^{\frac{1}{2}}}{\sqrt{\mathrm{Tr}\left[ E(m) \rho \right]}} E(m)^{\frac{1}{2}}S_u \rho^{\frac{1}{2}}\right]}^2
\nonumber \\
&\leqslant  \int dm~ \mathrm{Tr}\left[ E(m) S_u \rho S_u \right] = \mathrm{Tr}\left[ \rho S_u^2 \right] \equiv \mathcal{K}_u,
\label{quantuminequalities}
\end{align}
having used the Cauchy-Schwarz inequality $ |\mathrm{Tr}[X^\dagger Y]|^2 \leqslant \mathrm{Tr}[ X^\dagger X] \mathrm{Tr}[Y^\dagger Y]$ with $X = E(m)^{\frac{1}{2}} \rho^{\frac{1}{2}}/\lbrace\mathrm{Tr}\left[ E(m) \rho \right]\rbrace^{\frac{1}{2}}$, $Y = E(m)^{\frac{1}{2}} S_u \rho^{\frac{1}{2}}$. As we expected, Eq.\hspace{0.2em}(\ref{quantuminequalities}) is formally identical to the result by Braunstein and Caves \cite{BraunsteinCaves1994, genoni2008}.

Now we recall that $\theta_u = \sum_{i=1}^d u_i \theta_i$, implying that $\bar{\rho}_u = \sum_{i=1}^d u_i \bar{\rho}_i$, with $\bar{\rho}_i = \int d\boldsymbol{\theta} p(\boldsymbol{\theta}) \rho(\boldsymbol{\theta}) \theta_i$. In turn, this allows us to express $S_u$ as $S_u = \sum_{i=1}^d u_i S_i$, with $S_i \rho + \rho S_i = 2\bar{\rho}_i$ and $S_i$ Hermitian. Using these expressions and the fact that $\boldsymbol{u}$ is real, in Appendix \ref{matrixder} we show that $\mathcal{K}_u$ can be written explicitly as $\mathcal{K}_u = \boldsymbol{u}^\transpose \mathcal{K} \boldsymbol{u}$, where $\mathcal{K}_{ij} = \mathrm{Tr}\left[\rho \left(S_i S_j + S_j S_i \right) \right]/2$ are the components of $\mathcal{K}$.  

Given that the previous operations must be valid for any $\boldsymbol{u}$, we finally arrive at the chain of matrix inequalities
\begin{equation}
\Sigma_{\mathrm{mse}} \geqslant \Sigma_c \geqslant \Sigma_q = \int d\boldsymbol{\theta} p(\boldsymbol{\theta}) \boldsymbol{\theta}\boldsymbol{\theta}^\transpose - \mathcal{K}.
\label{mybound}
\end{equation}
The quantum bound is one of our central results. 

By combining now our inequality $\Sigma_{\mathrm{mse}} \geqslant \Sigma_q$ with the fact that $\mathcal{W}$ is positive semi-definite, we find that the single-shot uncertainty in Eq.\hspace{0.2em}(\ref{msegen}) is bounded as
\begin{equation}
\bar{\epsilon}_\mathrm{mse} \geqslant \sum_{i=1}^d w_i \left[\int d\boldsymbol{\theta} p(\boldsymbol{\theta})\theta_i^2 - \mathrm{Tr}\left(\rho S_i^2\right) \right].
\label{multibayesbound}
\end{equation}
In addition, using the identity $\mathrm{Tr}(\rho S_i) = \int d\boldsymbol{\theta}p(\boldsymbol{\theta})\theta_i$ \footnote{Note that, in the non-Bayesian theory, $\mathrm{Tr}[\rho(\boldsymbol{\theta})L_i(\boldsymbol{\theta})] = 0$.} we may rewrite Eq.\hspace{0.2em}(\ref{multibayesbound}) as the generalised uncertainty relation 
\begin{equation}
\bar{\epsilon}_\mathrm{mse} \geqslant \sum_{i=1}^d w_i \left(\Delta \theta_{p,i}^2 - \Delta S_{\rho, i}^2\right),
\end{equation}
where, for the $i$-th parameter, 
\begin{equation}
\Delta \theta_{p,i}^2 \equiv \int d\boldsymbol{\theta} p(\boldsymbol{\theta})\theta_i^2 - \left[\int d\boldsymbol{\theta} p(\boldsymbol{\theta})\theta_i \right]^2
\end{equation}
is the prior uncertainty and
\begin{equation}
\Delta S_{\rho, i}^2 \equiv \mathrm{Tr}(\rho S_i^2 ) - \mathrm{Tr}(\rho S_i )^2
\end{equation}
is an uncertainty associated with the quantum estimator.

%-------------------------------------------------------------------------

\subsection{Saturability conditions}\label{saturability}

As we have discussed, the classical result $\Sigma_{\mathrm{mse}} \geqslant \Sigma_c$ becomes an equality when the estimators are given by the averages over the posterior probability \cite{kay1993}, while the quantum bound $\Sigma_c \geqslant \Sigma_q$ relies on the inequalities in Eq.\hspace{0.2em}(\ref{quantuminequalities}). The first one is saturated when $\mathrm{Tr}\left[E(m) S_u \rho \right]$ is real, while the Cauchy-Schwarz inequality is saturated iff $X \propto Y$ \cite{helstrom1968multiparameter}, which implies that $E(m)^{\frac{1}{2}}\rho^{\frac{1}{2}}/\mathrm{Tr}\left[E(m) \rho \right] = E(m)^{\frac{1}{2}}S_u\rho^{\frac{1}{2}}/\mathrm{Tr}\left[E(m) S_u \rho \right]$. 

If $[S_i, S_j] = 0$ for all $i$, $j$, then we may fulfil such conditions by constructing the measurement with the projections onto the common eigenstates of this set of commuting operators. To verify it, let us rewrite $S_u$ as $S_u = \int dm~c_u(m) \ketbra{\psi(m)}$, 
where $c_u(m) = \sum_{i=1}^d u_i c_i(m)$, $\lbrace c_i(m) \rbrace$ are the eigenvalues of $S_i$ and $\lbrace \ketbra{\psi(m)} \rbrace$ are the common eigenstates of $\lbrace S_i \rbrace$. Then, by using $E(m) =  \ketbra{\psi(m)}$ we find the required result
\begin{eqnarray}
B_u &=& \int dm\left(\frac{\mathrm{Re}\left\lbrace\mathrm{Tr}\left[\ketbra{\psi(m)} S_u\rho\right]\right\rbrace}{\sqrt{\mathrm{Tr}\left[ \ketbra{\psi(m)} \rho\right]}}\right)^2
\nonumber \\
&=& \int dm~ c_u^2(m) \mathrm{Tr}\left[\ketbra{\psi(m)} \rho\right] = \mathcal{K}_u.
\end{eqnarray}

Unfortunately, it is known that the optimal strategy for Bayesian multi-parameter estimation is not necessarily based on the projective measurements that are independently optimal \cite{helstrom1974, personick1969thesis}, which reflects the fact that the operators $\lbrace S_i \rbrace$ might not commute. In fact, the optimal strategy generally requires generalised measurements \cite{helstrom1974}. Thus our bound cannot always be saturated.

\footnotetext[2020]{An explicit example with NOON states where the single-parameter version of Eq.\hspace{0.2em}(\ref{multibayesbound}) is compared with other Bayesian bounds can be found in Sec. 5.4 of \cite{jesus2019thesis}.}

\footnotetext[2021]{We leave for future work to examine the relative tightness of our bound with respect to alternatives such as the multi-parameter Weiss-Weinstein bound \cite{tsang2016} or the Yuen-Lax bound for complex quantities \cite{yuen1973} when the latter is applied to real parameters.}

Despite this, we will show that this bound can still be useful and informative. Indeed, the results based on it are tight and fundamental whenever the operators $\lbrace S_i \rbrace$ commute, and the complexity of its calculation is similar to that of the Fisher information matrix for density matrices \footnote{Note that this type of calculation can be very challenging when the dimension of the space is large (see, e.g., \cite{tsang2019}).}, with the extra advantage of not having to invert $\mathcal{K}$. Furthermore, any other multi-parameter bound for Eq.\hspace{0.2em}(\ref{msegen}) that also ignores the potential non-commutativity of $\lbrace S_i \rbrace$ will necessarily be equal to or lower than Eq.\hspace{0.2em}(\ref{multibayesbound}), since the quantity $\int d\boldsymbol{\theta} p(\boldsymbol{\theta})\theta_i^2 - \mathrm{Tr}\left(\rho S_i^2\right)$ is the optimum for the estimation of $\theta_i$ \cite{personick1971, yuen1973, helstrom1976}. Therefore, our result will produce tighter bounds than proposals such as the multi-parameter quantum Ziv-Zakai bound \cite{zhang2014, Note2020, Note2021}.

\subsection{Extension to several repetitions}\label{manyrep}

For $\mu$ identical and independent trials the likelihood becomes $p(\boldsymbol{m}|\boldsymbol{\theta}) = \prod_{i=1}^\mu \mathrm{Tr}[E(m_i) \rho(\boldsymbol{\theta})]$, where $m_i$ is the outcome of the $i$-th iteration. Using $p(\boldsymbol{\theta}, \boldsymbol{m}) = p(\boldsymbol{\theta}) p(\boldsymbol{m}|\boldsymbol{\theta})$, the uncertainty including the information from all the repetitions is
\begin{equation}
\bar{\epsilon}_{\mathrm{mse}} = \sum_{i=1}^d w_i \int d\boldsymbol{\theta} d\boldsymbol{m} ~p(\boldsymbol{\theta}, \boldsymbol{m}) \left[g_i(\boldsymbol{m}) - \theta_i  \right]^2.
\label{msegenmany}
\end{equation}

To generalise our single-parameter methodology in \cite{jesus2018} we just need to calculate Eq.\hspace{0.2em}(\ref{msegenmany}) numerically \footnote{Here we use our numerical algorithm for multi-parameter metrology in \cite{jesus2019thesis}.} after selecting the optimal estimators $\boldsymbol{g}(\boldsymbol{m}) = \int d\boldsymbol{\theta} p(\boldsymbol{\theta}|\boldsymbol{m}) \boldsymbol{\theta}$ and the optimal single-shot measurement $E(m_i)=\ketbra{\psi(m_i)}$, provided that the latter exists.  

%-------------------------------------------------------------------------

\section{Applications}\label{app}

\subsection{Qubit sensing network}\label{qubitsec}

Our first example is a qubit network prepared as $\ket{\psi_0} = [\ket{00}+\gamma (\ket{01}+\ket{10})+\ket{11}]/\sqrt{2(1+\gamma^2)}$, with real $\gamma$, which upon interacting with the portion of environment that we wish to sense is transformed by $U(\theta_1, \theta_2) = \mathrm{exp}[-i(\sigma_{z,1}\theta_1 + \sigma_{z,2}\theta_2)/2]$, where $\sigma_{z,1} = \sigma_z \otimes \mathbb{I}$, $\sigma_{z,2} =  \mathbb{I}\otimes \sigma_z$, $\sigma_z$ is a Pauli matrix and $\mathbb{I}$ is the identity matrix. Furthermore, we assume equally important parameters (i.e., $\mathcal{W} = \mathbb{I}/2$). This sensing network was proposed and studied in \cite{proctor2017networked} using the quantum Cram\'{e}r-Rao bound, and the latter was found to be $\bar{\epsilon}_{\mathrm{cr}}= \mathrm{Tr}(\mathcal{W} F_q^{-1})/\mu = (1+\gamma^2)^2/(4\mu\gamma^2)$, where $F_q$ is the quantum Fisher information matrix \footnote{The components of the quantum Fisher information matrix for pure states and  $U(\boldsymbol{\theta}) = \mathrm{exp}(-i \boldsymbol{K}\cdot \boldsymbol{\theta})$ are $(F_q)_{ij} =  4 ( \expval{K_i K_j}{\psi_0} - \expval{K_i}{\psi_0}\expval{K_j}{\psi_0} )$ \cite{proctor2017networked}.}.

\begin{figure}[t]
\centering
\includegraphics[trim={0.1cm 0.1cm 1cm 0.5cm},clip,width=8.6cm]{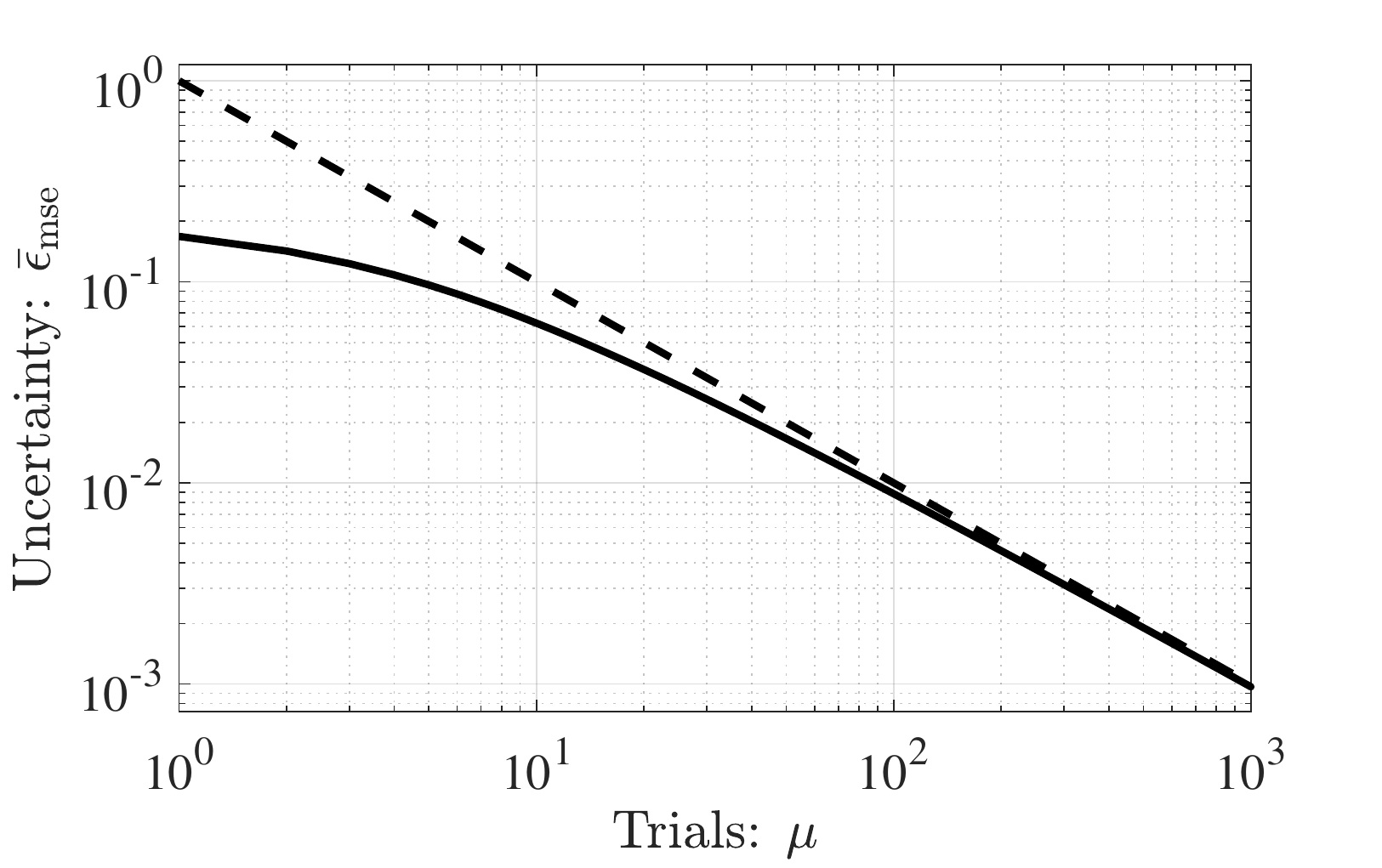}
	\caption{Mean square error optimised shot by shot (solid line) and quantum Cram\'{e}r-Rao bound (dashed line) for the qubit network in Sec.\hspace{0.2em}\ref{qubitsec}, with $\gamma = 1$ and a prior area $\pi^2/4$ centred around $(0, 0)$. This scheme is optimal at least for $\mu = 1$ and $\mu \gg 1$. Note that $\bar{\epsilon}_{\mathrm{cr}} = 1/\mu$ when $\gamma = 1$ \cite{proctor2017networked}.}
\label{multibayes_plot}
\end{figure}

Let us start with the single-shot analysis. Assuming moderate prior knowledge given by $p(\theta_1, \theta_2) = 4/\pi^2$ when $(\theta_1, \theta_2) \in [-\pi/4, \pi/4]\times[-\pi/4, \pi/4]$, and zero otherwise, the quantum estimators arising from $S_i \rho + \rho S_i = 2\bar{\rho_i}$, $\rho = \int d\boldsymbol{\theta} p(\boldsymbol{\theta})\rho(\boldsymbol{\theta})$ and $\bar{\rho}_i = \int d\boldsymbol{\theta} p(\boldsymbol{\theta})\rho(\boldsymbol{\theta})\theta_i$ are 
\begin{equation}
S_1 = \frac{2 \left(4-\pi\right)}{\pi\left(1+\gamma^2\right)}\left[ \frac{\gamma}{\sqrt{2}} \sigma_y\otimes\mathbb{I} + \frac{1-\gamma^2}{\pi}\sigma_x\otimes\sigma_y  \right],
\label{qest1}
\end{equation}
\begin{equation}
S_2 = \frac{2 \left(4-\pi\right)}{\pi\left(1+\gamma^2\right)}\left[ \frac{\gamma}{\sqrt{2}} \mathbb{I}\otimes\sigma_y + \frac{1-\gamma^2}{\pi}\sigma_y\otimes\sigma_x  \right],
\label{qest2}
\end{equation}
where the columns are labelled as $\ket{00}$, $\ket{01}$, $\ket{10}$ and $\ket{11}$,  and $\sigma_x$ and $\sigma_y$ are Pauli matrices. In addition, the bound in Eq.\hspace{0.2em}(\ref{multibayesbound}) implies that
\begin{eqnarray}
\bar{\epsilon}_{\mathrm{mse}} \geqslant \frac{\pi^2}{48} - \frac{2\left(4-\pi\right)^2\left[2-\left(4-\pi^2\right)\gamma^2 + 2\gamma^4 \right]}{\pi^4 \left(1+\gamma^2\right)^2}
\label{qnetworksingleshot}
\end{eqnarray}
for $\mu = 1$, which achieves its minimum at $\gamma = \pm 1$. That is, $\bar{\epsilon}_{\mathrm{mse}} \geqslant \pi^2/48 - (4-\pi)^2/(2\pi^2) \approx 0.168$. Appendix\hspace{0.2em}\ref{qubitapp} provides the details of these calculations.

Since $[S_1, S_2] = 0$, there is a measurement achieving the minimum single-shot error above. Choosing $\gamma = 1$ we can construct an optimal strategy given by the projectors $\ket{s_+, s_+}$, $\ket{s_-, s_-}$, $\ket{s_+, s_-}$, $\ket{s_-, s_+}$, where $\ket{s_\pm} = (\ket{0}\pm i\ket{1})/\sqrt{2}$, and calculate the uncertainty for $\mu$ trials in Eq.\hspace{0.2em}(\ref{msegenmany}) using this measurement in each shot. The solid line in Fig.\hspace{0.2em}\ref{multibayes_plot} shows this numerical result, while the dashed line is the quantum Cram\'{e}r-Rao bound. Given that the latter is approached by the Bayesian error as $\mu$ grows, we see that our Bayesian strategy is optimal both for $\mu = 1$ and $\mu \gg 1$, in consistency with the behaviour observed in the single-parameter case \cite{jesus2018}. 

Additionally, note that the asymptotic theory is a good approximation to our bound (i.e., their relative error is less than $5\%$ \cite{jesus2017, jesus2019thesis}) only when $\mu > 5.05 \cdot 10^2$, while there exist practical multi-parameter schemes where, e.g.,  $\mu = 3\cdot 10^2$ \cite{berchera2019} and $\mu = 10^2$ \cite{polino2018}. This demonstrates the potential relevance of our approach in experiments. 

In fact, it can be shown that our optimisation of this protocol provides a meaningful amount of information even for a single shot. Indeed, by noticing that $\bar{\epsilon}_{\mathrm{prior}} = \pi^2/48 \approx 0.206$, and defining a notion of improvement as $(\bar{\epsilon}_{\mathrm{prior}}-\bar{\epsilon}_{\mathrm{mse}})/\bar{\epsilon}_{\mathrm{prior}}$ multiplied by $100\%$, we see that a single shot improves our knowledge about $(\theta_1, \theta_2)$ by $18\%$ with respect to the prior uncertainty \footnote{Further examples of this notion of improvement can be found in \cite{jesus2018dec, jesus2019thesis}}. 

From a fundamental perspective, a remarkable conclusion of our analysis is that this qubit sensing network reaches its optimal single-shot error without entanglement, since the strategy above is \emph{local} in the sense that both the state and the measurement are separable. In other words, we have shown that this result, which previously had been established only in an asymptotic fashion \cite{proctor2017networked}, is also valid for repeated experiments with limited data and moderate prior.
%-------------------------------------------------------------------------

\subsection{Quantum imaging}

Secondly we wish to examine the phase imaging model explored in \cite{humphreys2013, knott2016local} with the Cram\'{e}r-Rao bound, and in \cite{chiara2003} using covariant measurements. In the former the scheme operates asymptotically, while the latter assumes no prior knowledge. On the contrary, our protocol will assume an intermediate amount of prior information.

Consider a system with $(d+1)$ optical modes where we encode a phase shift $\theta_j$ with a local unitary $U(\theta_j) = \mathrm{exp}(-i a_j^\dagger a_j \theta_j)$ in the $j$-th mode, for $1\leqslant j\leqslant d$, while $j = 0$ is a reference mode calibrated in advance \cite{proctor2017networked}. Note that $a_j^\dagger$ and $a_j$ are creation and annihilation operators. Given this, one possibility is following a \emph{global} approach and preparing the entangled probe \cite{knott2016local, humphreys2013}
\begin{eqnarray}
\ket{\psi_0} = \frac{1}{\sqrt{d + \alpha^2}}\left(\alpha \ket{\bar{n}_0} + \sum_{k=1}^d \ket{\bar{n}_k}\right),
\label{gennoonstate}
\end{eqnarray}
where $\ket{\bar{n}_j}$ is a state with $\bar{n}$ photons in the $j$-th mode and zero in the rest, $\bar{n}$ is the mean number of quanta and $\alpha$ is assumed to be real. This is a generalised NOON state.

Suppose that the unknown parameters are equally important, so that $\mathcal{W} = \mathbb{I}/d$, and consider a flat prior of hypervolume $(2\pi/\bar{n})^d$ with $\bar{n} \geqslant 4$ and centred around $(0, 0, \dots)$. This  prior knowledge is sufficient to avoid the periodicities associated with NOON states and to employ the square error in phase estimation \cite{jesus2017, jesus2018, alfredo2017, hall2012, friis2017}.  

Our calculations in Appendix\hspace{0.2em}\ref{imaging1} show that, for this scheme, 
\begin{equation}
S_k =  \frac{- 2i\alpha}{\bar{n}\left(1 + \alpha^2\right)} \left(\ketbra{\bar{n}_k}{\bar{n}_0} - \ketbra{\bar{n}_0}{\bar{n}_k}\right)
\label{qestimaging}
\end{equation} 
after solving $S_k \rho + \rho S_k = 2\bar{\rho_k}$, $\rho = \int d\boldsymbol{\theta} p(\boldsymbol{\theta})\rho(\boldsymbol{\theta})$ and $\bar{\rho}_k = \int d\boldsymbol{\theta}p(\boldsymbol{\theta})\rho(\boldsymbol{\theta})\theta_k$, and that Eq.\hspace{0.2em}(\ref{multibayesbound}) becomes
\begin{eqnarray}
\bar{\epsilon}_{\mathrm{mse}} \geqslant \frac{1}{\bar{n}^2} \left[\frac{\pi^2}{3} - \frac{4 \alpha^2}{\left(1 + \alpha^2\right)\left(d + \alpha^2\right)} \right].
\label{globalboundfree}
\end{eqnarray}
Since the latter achieves its minimum at $\alpha = d^{1/4}$,
\begin{equation}
\bar{\epsilon}_{\mathrm{mse}} \geqslant \frac{1}{\bar{n}^2}\left[\frac{\pi^2}{3} - \frac{4}{(1+\sqrt{d})^2} \right] ~\underset{d\gg 1}{\longrightarrow}~ \frac{1}{\bar{n}^2}\left(\frac{\pi^2}{3} - \frac{4}{d}\right)
\label{globalscaling}
\end{equation}
is the single-shot bound for the global scheme.

Unlike in the qubit case, here $[S_k, S_j] \neq 0$ when $k \neq j$, which means that we cannot extract an optimal measurement as we did before. However, the asymptotic theory has been shown to predict that the scaling associated with the state in Eq.\hspace{0.2em}(\ref{gennoonstate}) can also be achieved with a local scheme \cite{knott2016local}. Therefore, if we could recover this phenomenon using our Bayesian bound, then we may be able to construct a single-shot strategy with a precision that scales as in Eq.\hspace{0.2em}(\ref{globalscaling}).

In a local protocol we have that $\rho_0 = \rho_0^{\mathrm{ref}}\otimes \rho_0^{(1)}\otimes \cdots  \otimes \rho_0^{(d)}$, with $\rho_0^{(k)} = |\phi_0^{(k)}\rangle \langle \phi_0^{(k)}|$ for pure states. Following \cite{knott2016local} we choose $\ket{\phi_0}$ as 
\begin{equation}
\ket{\phi_0} = \left[\sqrt{1- \frac{\bar{n}}{N(d+1)}}\ket{0}+\sqrt{\frac{\bar{n}}{N(d+1)}}\ket{N}\right],
\label{localstrategy}
\end{equation}
where $N$ can vary while $\bar{n}$ remains constant. This is an important property, since it allows us to modify the precision through $N$ without altering the average amount of resources. More concretely, inserting the resulting state in Eq.\hspace{0.2em}(\ref{multibayesbound}) we find the bound (see Appendix \ref{imaging1})
\begin{equation}
\bar{\epsilon}_{\mathrm{mse}}\geqslant \left[\pi^2/3 - f(N,\bar{n},d)\right]/\bar{n}^2,
\end{equation}
where $f(N,\bar{n},d)$ satisfies that: 
\begin{enumerate}
\item[a)] if $N = \bar{n}$, then $f(N,\bar{n},d) = 4d/(1+d)^2$, and
\begin{equation}
\bar{\epsilon}_{\mathrm{mse}} \geqslant \frac{1}{\bar{n}^2}\left[\frac{\pi^2}{3} - \frac{4d}{(1+d)^2}\right] ~~\underset{d\gg 1}{\longrightarrow}~~ \frac{1}{\bar{n}^2}\left(\frac{\pi^2}{3} - \frac{4}{d}\right);
\label{localscaling}
\end{equation}
\item[b)] if $N \rightarrow \infty$, then $f(N,\bar{n},d) \rightarrow 0$, so that
\begin{equation}
\bar{\epsilon}_\mathrm{mse}  ~~\underset{N \rightarrow \infty}{\longrightarrow}~~  \frac{\pi^2}{3 \bar{n}^2} = \frac{1}{d} \sum_{i=1}^d \Delta \theta_{p,i}^2.
\label{firstproperty}
\end{equation}
\end{enumerate}

The scaling in Eq.\hspace{0.2em}(\ref{localscaling}) is exactly that found in Eq.\hspace{0.2em}(\ref{globalscaling}) for the global strategy, and the bound associated with the local scheme can be reached. To see how, first note that if the parameters are a priori thought of as independent (a condition fulfilled by our separable prior), then $\rho = \rho_0^{\mathrm{ref}}\otimes \rho^{(1)}\otimes  \cdots \otimes \rho^{(d)}$ and $\bar{\rho}_k = \rho_0^{\mathrm{ref}}\otimes \rho^{(1)}\otimes  \cdots \otimes \bar{\rho}^{(k)} \otimes  \cdots \otimes \rho^{(d)}$. In turn, $S_k =  \mathbb{I}_{\mathrm{ref}}\otimes\mathbb{I}\otimes \cdots \otimes S^{(k)} \otimes\cdots \otimes \mathbb{I}$, so that $S_k$ commutes trivially with the rest. Hence, despite the non-commutative nature of the estimators associated with the original global protocol, we can still construct a strategy with the desired scaling by using local states and measurements, as we did with the qubit network.

The previous discussion also shows that, for protocols with moderate prior knowledge and limited measurement data, a global strategy \emph{is not required} to achieve the scaling predicted in Eq.\hspace{0.2em}(\ref{globalscaling}) \footnote{Although other global strategies might still offer a better precision}. This is similar to our conclusion in Sec.\hspace{0.2em}\ref{qubitsec} for the qubit network, but the imaging protocol provides a more fundamental way of understanding the importance of this result.

First we recall that, according to the asymptotic theory, the precision of schemes such as Eq.\hspace{0.2em}(\ref{localstrategy}) appears to grow unbounded as $N$ is increased \cite{rivas2012, knott2016local, jesus2017, tsang2012, lee2019}, while a finite precision is known to be recovered when the prior information is appropriately taken into account \cite{tsang2012, berry2012infinite, hall2012, jesus2017, giovannetti2012subheisenberg, pezze2013, rafal2015, jesus2019thesis}. This means that our local scheme will not be able to produce an arbitrarily good precision by simply increasing $N$, and this is precisely what Eq.\hspace{0.2em}(\ref{firstproperty}) demonstrates. To understand it, note that the periodicity in Eq.\hspace{0.2em}(\ref{localstrategy}) is $2\pi/N$, so that the width where the value of each phase may lie needs to be smaller as $N$ grows to avoid ambiguities, and thus the limit $N \rightarrow \infty$ is equivalent to requiring that the parameters are practically localised a priori. Since the prior knowledge is fixed by the situation under analysis, eventually the high amount of prior information needed as $N$ grows is not provided, and the scheme is unable to extract more information beyond our initial knowledge.

Therefore, if we were to restrict our calculations to the framework provided by the quantum Fisher information matrix and the associated Cram\'{e}r-Rao bound, then one could question the physical validity of searching for local strategies that are as sensitive as a global scheme (see, e.g., \cite{knott2016local, proctor2017networked, proctor2017networkedshort, altenburg2018}), for the apparent enhancement of the local approach might not be such. Our Bayesian analysis shows that this type of result can in fact emerge in a more realistic regime with finite resources and without involving unbounded precisions, and this puts the idea that some local schemes reproduce the enhancements predicted by certain global protocols on a solid basis.

%-------------------------------------------------------------------------

\section{Discussion and conclusions}\label{conclusions}

Our method  offers a powerful and novel framework to study schemes with limited data and moderate prior knowledge, a regime of practical interest and often outside of the scope of existing techniques. Given that experimental multi-parameter protocols are already a reality \cite{roccia2018, polino2018, berchera2019, valeri2020}, our proposal could play a crucial role in the design of future experiments.

Theoretically, a major strength of our derivation of Eq.\hspace{0.2em}(\ref{multibayesbound}) is its clear separation of the classical optimisation from the quantum problem, in analogy with the proof of the Braunstein-Caves inequality \cite{BraunsteinCaves1994, genoni2008}. One could be tempted to argue that by introducing $S_u \rho + \rho S_u = 2 \bar{\rho}_u$ we are somehow assuming the answer, as this is the solution of the single-parameter optimisation. However, here this equation is a redefinition of $\bar{\rho}_u$ allowing us to derive a bound, and its form is imposed by the formal analogy with the Fisher information. Moreover, given the scalar quantity $\boldsymbol{u}^\transpose \Sigma_{\mathrm{mse}}\boldsymbol{u}$, we could instead employ any of the alternative single-parameter proofs available \cite{personick1971, helstrom1976, macieszczak2014bayesian} to show that $\boldsymbol{u}^\transpose \Sigma_{\mathrm{mse}}\boldsymbol{u} \geqslant \int d\boldsymbol{\theta}p(\boldsymbol{\theta})\theta_u^2 - \mathrm{Tr}(\rho S_u^2)$, from where (\ref{multibayesbound}) follows, although these approaches, unlike ours, merge classical and quantum steps. 

Among all the Bayesian bounds neglecting the interference between individually optimal quantum strategies, our result is to be preferred, since it recovers the true optimum both in the single-parameter limit and when $\lbrace S_i \rbrace$ commute. Moreover, our qubit and optical examples demonstrate that its calculation can be tractable, and that the Bayesian nature of our approach can produce a more physical picture of the performance associated with multi-parameter protocols. Additionally, note that while Eq.\hspace{0.2em}(\ref{multibayesbound}) may not always produce tight bounds, we can still use it to study how close a given measurement can get (see Appendix\hspace{0.2em}\ref{imaging2} for an example of this type of calculation). Hence, this tool will be very useful to enquire about fundamental limits or the precision scaling in a range of practical cases. 

Finally, our approach could be key to finding fundamental bounds including the case of non-commuting estimators. In particular, the rationale behind our Bayesian analogue of the Helstrom Cram\'{e}r-Rao bound might also lead us to a similar Bayesian analogue of the Holevo Cram\'{e}r-Rao bound \footnote{Some results in this direction can be found in \cite{gill2011, rafal2020, gill2020}.}, which could eliminate the deficiencies of both our result and the standard Holevo Cram\'{e}r-Rao bound and move us closer to the optima predicted by Holevo and Helstrom's often intractable fundamental equations \cite{helstrom1976, helstrom1974, holevo1973b, holevo1973, jesus2019thesis}. This path promises a bright future for multi-parameter metrology. 

%-------------------------------------------------------------------------

\begin{acknowledgments}
J.R. thanks Francesco Albarelli for helpful comments at the CEWQO 2019 about the saturability of different multi-parameter bounds of an asymptotic or local nature. We also thank Jasminder Sidhu, Mankei Tsang, Mark Bason and Nathan Babcock for helpful discussions. This work was funded by the South East Physics Network (SEPnet) and the United Kingdom EPSRC through the Quantum Technology Hub: Networked Quantum Information Technology (grant reference EP/M013243/1). J.R. also acknowledges support from Engineering and Physical Sciences Research Council (UKRI) grant EP/T002875/1.
\end{acknowledgments}

% References
\bibliography{references_12022020}

\appendix
%\onecolumngrid

\section{Optimal classical estimators}\label{classicalest}

The form of the optimal classical estimators for the multi-parameter square error is a well-known result \cite{kay1993}. To recover it as the matrix inequality $\Sigma_\mathrm{mse} \geqslant \Sigma_c$, first we note that the variational problem
\begin{equation}
\delta \epsilon\left[g_u(m)\right] = \delta \int dm~\mathcal{L}\left[m, g_u(m) \right] = 0,
\label{variationalprobsupp}
\end{equation} 
with $\mathcal{L}\left[m, g_u(m) \right] = \int d\boldsymbol{\theta} p(\boldsymbol{\theta}, m) \left[g_u(m) - \theta_u  \right]^2$ and $x_u = \sum_{i=1}^d u_i x_i$, is equivalent to require that \cite{mathematics2004}
\begin{equation}
\frac{d}{d\beta} \epsilon\left[g_u(m)+\beta h_u(m)\right] \bigg\rvert_{\beta = 0} = 0,~~\text{for~all}~~h_u(m). 
\end{equation}
In our case we have that
\begin{align}
\label{firstvariation}
\frac{d}{d\beta}  &\epsilon\left[g_u(m)+\beta h(m)\right] 
\\ 
&= 2 \int d\boldsymbol{\theta} dm~p(\boldsymbol{\theta}, m) \left[g_u(m) + \beta h_u(m) - \theta_u \right] h_u(m),
\nonumber
\end{align}
which means that the requirement to find the extrema is
\begin{align}
\frac{d}{d\beta} &\epsilon\left[g_u(m) + \beta h_u(m)\right] \bigg\rvert_{\beta = 0} 
\nonumber \\
&= 2 \int d\boldsymbol{\theta} dm~p(\boldsymbol{\theta}, m) \left[ g_u(m) - \theta_u \right] h_u(m) = 0,
\label{variationcondition}
\end{align}
and this implies that $\int d\boldsymbol{\theta} p(\boldsymbol{\theta}, m) [g_u(m) - \theta_u] = 0$ if Eq.\hspace{0.2em}(\ref{variationcondition}) is to be satisfied by an arbitrary $h_u(m)$. By decomposing the joint probability as $p(\boldsymbol{\theta}, m) = p(m) p(\boldsymbol{\theta}|m)$, with $p(\boldsymbol{\theta}|m) \propto p(\boldsymbol{\theta})p(m|\boldsymbol{\theta})$, we see that the solution
\begin{equation}
g_u(m) = \int d\boldsymbol{\theta} p(\boldsymbol{\theta}|m) \theta_u
\end{equation}
makes the error $\epsilon\left[g_u(m)\right]$ extremal.

To verify that this is a minimum we can use the functional version of the second derivative test. Calculating the second variation from Eq.\hspace{0.2em}(\ref{firstvariation}) we see that
\begin{align}
\frac{d^2}{d\beta^2}  &\epsilon\left[g_u(m)+\beta h_u(m)\right]\bigg\rvert_{\beta = 0} 
\nonumber \\
&= 2 \int d\boldsymbol{\theta} dm~p(\boldsymbol{\theta}, m) h_u(m)^2 > 0
\end{align}
for non-trivial variations. Thus $g_u(m) = \int d\theta p(\boldsymbol{\theta}|m) \theta_u$ gives the minimum of $\epsilon\left[g_u(m)\right] = \boldsymbol{u}^\transpose \Sigma_{\mathrm{mse}} \boldsymbol{u}$, and 
\begin{align}
\boldsymbol{u}^\transpose \Sigma_{\mathrm{mse}} \boldsymbol{u} &\geqslant \int \hspace{-0.15em} d\boldsymbol{\theta} p(\boldsymbol{\theta})\theta_u^2
- \int dm  \frac{\left[ \int d\boldsymbol{\theta} p(\boldsymbol{\theta})p(m|\boldsymbol{\theta}) \theta_u \right]^2}{\int d\boldsymbol{\theta} p(\boldsymbol{\theta})p(m|\boldsymbol{\theta})} 
\nonumber \\
&= \boldsymbol{u}^\transpose \Sigma_c \boldsymbol{u}
\end{align} 
for any $\boldsymbol{u}$, where $\Sigma_c$ is the optimal matrix error
\begin{align}
\Sigma_c =& \int d\boldsymbol{\theta} p(\boldsymbol{\theta})\boldsymbol{\theta}\boldsymbol{\theta}^\transpose
 \\
&- \int dm  \frac{\left[ \int d\boldsymbol{\theta} p(\boldsymbol{\theta})p(m|\boldsymbol{\theta}) \boldsymbol{\theta} \right]\left[ \int d\boldsymbol{\theta} p(\boldsymbol{\theta})p(m|\boldsymbol{\theta}) \boldsymbol{\theta}\right]^\transpose}{\int d\boldsymbol{\theta} p(\boldsymbol{\theta})p(m|\boldsymbol{\theta})}.
\nonumber
\end{align}
In this way we have arrived at the desired matrix inequality, i.e.,  $\Sigma_{\mathrm{mse}} \geqslant \Sigma_c$.

\section{Matrix form of $\mathcal{K}_u$}\label{matrixder}

A crucial step to find the quantum inequality $\Sigma_c \geqslant \Sigma_q$ that constitutes one of our main results is to rewrite $\mathcal{K}_u = \mathrm{Tr}(\rho S_u^2)$ as $\mathcal{K}_u = \boldsymbol{u}^\transpose \mathcal{K} \boldsymbol{u}$, where $\mathcal{K}_{ij} = \mathrm{Tr}(\rho A_{ij})$ is a matrix and $A_{ij}$ is some operator associated with the product of $S_i$ and $S_j$. Since $S_i$ and $S_j$ might not commute, in principle we could consider either $A_{ij} = S_i S_j$, $A_{ij} = S_j S_i$ or $A_{ij} = (S_i S_j + S_j S_i)/2$. However, if we first decompose $A_{ij}$ as $2 A_{ij} = (A_{ij}+A_{ij}^\dagger) + (A_{ij} - A_{ij}^\dagger)$, so that
\begin{equation}
\boldsymbol{u}^\transpose \mathcal{K} \boldsymbol{u} 
= \frac{1}{2}\left(\sum_{i,j=1}^d u_i u_j C^{+} + \sum_{i,j=1}^d u_i u_j C^{-}\right),
\end{equation}
with $C^{\pm} = \mathrm{Tr}[\rho(A_{ij}  \pm  A_{ij}^\dagger)]$, and observing that $\boldsymbol{u}$ is real and $S_i$ Hermitian, then
\begin{align}
\boldsymbol{u}^\transpose \mathcal{K} \boldsymbol{u} &= \frac{1}{2}\sum_{i, j=1}^d u_i u_j\mathrm{Tr}\left[\rho\left(S_i S_j +  S_j S_i\right)\right] 
\nonumber \\
&= \mathrm{Tr}\left(\rho S_u^2\right) = \mathcal{K}_u
\end{align}
for any of the three possible forms of $A_{ij}$. Therefore, we can take $\mathcal{K}$ to be a symmetric matrix with elements
\begin{equation}
\mathcal{K}_{ij} = \mathrm{Tr}\left[\rho \left(S_i S_j + S_j S_i \right) \right]/2.
\label{bayesinfmatrix}
\end{equation}
See \cite{helstrom1968multiparameter} for the analogous step in the derivation of the multi-parameter quantum Cram\'{e}r-Rao bound.

\section{Calculations for the qubit network}\label{qubitapp}

Given the network state
\begin{equation}
\ket{\psi_0} =\frac{1}{\sqrt{2(1+\gamma^2)}}\left[\ket{00}+\gamma (\ket{01}+\ket{10})+\ket{11}\right],
\end{equation}
the unitary encoding
\begin{equation}
U(\theta_1, \theta_2) = \mathrm{exp}\left[-i \left(\sigma_{z,1}\theta_1 + \sigma_{z,2}\theta_2\right)/2\right],
\end{equation}
and the prior density $p(\theta_1, \theta_2) = 4/\pi^2$, when $(\theta_1, \theta_2) \in [-\pi/4, \pi/4]\times[-\pi/4, \pi/4]$, we have that
\begin{align}
\rho &= \int d\theta_1 d\theta_2 p(\theta_1, \theta_2) \rho(\theta_1, \theta_2) 
\nonumber \\
&= \frac{4}{\pi^2}\int_{-\pi/4}^{\pi/4} d\theta_1 \int_{-\pi/4}^{\pi/4} d\theta_2\hspace{0.15em} U(\theta_1, \theta_2)\ketbra{\psi_0}U^{\dagger}\hspace{-0.1em}(\theta_1, \theta_2)
\nonumber \\
&= \frac{1}{2\pi^2(1+\gamma^2)}
\left(
\begin{array}{cccc}
 \pi ^2 & 2 \sqrt{2} \pi  \gamma  & 2 \sqrt{2} \pi  \gamma  & 8 \\
 2 \sqrt{2} \pi  \gamma  & \pi ^2 \gamma ^2 & 8 \gamma ^2 & 2 \sqrt{2} \pi  \gamma  \\
 2 \sqrt{2} \pi  \gamma  & 8 \gamma ^2 & \pi ^2 \gamma ^2 & 2 \sqrt{2} \pi  \gamma  \\
 8 & 2 \sqrt{2} \pi  \gamma  & 2 \sqrt{2} \pi  \gamma  & \pi ^2 \\
\end{array}
\right),
\label{labeleffstate}
\end{align}
\begin{align}
\bar{\rho}_1 &= \int d\theta_1 d\theta_2 p(\theta_1, \theta_2) \rho(\theta_1, \theta_2)\theta_1 
\nonumber \\
&= \frac{4}{\pi^2}\int_{-\pi/4}^{\pi/4} d\theta_1 \int_{-\pi/4}^{\pi/4} d\theta_2\hspace{0.15em} U(\theta_1, \theta_2)\ketbra{\psi_0}U^{\dagger}\hspace{-0.1em}(\theta_1, \theta_2)\theta_1
\nonumber \\
&= \frac{i(4-\pi)}{2\sqrt{2}\pi^2(1+\gamma^2)}
\begin{pmatrix}
0 & 0 & -\pi\gamma & -2\sqrt{2} \\
0 & 0 & -2\sqrt{2}\gamma^2 & -\pi\gamma \\
\pi\gamma & 2\sqrt{2}\gamma^2 & 0 & 0 \\
2\sqrt{2} & \pi\gamma & 0 & 0 
\end{pmatrix},
\label{labeleffmean1}
\end{align}
and
\begin{align}
\bar{\rho}_2 &= \int d\theta_1 d\theta_2 p(\theta_1, \theta_2) \rho(\theta_1, \theta_2)\theta_2 
\nonumber \\
&= \frac{4}{\pi^2}\int_{-\pi/4}^{\pi/4} d\theta_1 \int_{-\pi/4}^{\pi/4} d\theta_2\hspace{0.15em} U(\theta_1, \theta_2)\ketbra{\psi_0}U^{\dagger}\hspace{-0.1em}(\theta_1, \theta_2)\theta_2
\nonumber \\
&= \frac{i(4-\pi)}{2\sqrt{2}\pi^2(1+\gamma^2)}
\begin{pmatrix}
0 & -\pi\gamma & 0 & -2\sqrt{2} \\
\pi\gamma & 0 & 2\sqrt{2}\gamma^2 & 0 \\
0 & -2\sqrt{2}\gamma^2 & 0 & -\pi\gamma \\
2\sqrt{2} & 0 & \pi\gamma & 0 
\end{pmatrix},
\label{labeleffmean2}
\end{align}
where the columns are labelled as $\ket{00}$, $\ket{01}$, $\ket{10}$ and $\ket{11}$. In addition, by inserting Eqs.\hspace{0.2em}(\ref{labeleffstate} - \ref{labeleffmean2}) in $S_i\rho + \rho S_i = 2\bar{\rho}_i$ and using a computing system such as Mathematica to solve this Sylvester equation, we find that
\begin{align}
S_1 &= \frac{2 \left(4-\pi\right)}{\pi\left(1+\gamma^2\right)}\left( \frac{\gamma}{\sqrt{2}} \sigma_y\otimes\mathbb{I} + \frac{1-\gamma^2}{\pi}\sigma_x\otimes\sigma_y \right),
\nonumber \\
S_2 &= \frac{2 \left(4-\pi\right)}{\pi\left(1+\gamma^2\right)}\left( \frac{\gamma}{\sqrt{2}} \mathbb{I}\otimes\sigma_y + \frac{1-\gamma^2}{\pi}\sigma_y\otimes\sigma_x \right),
\label{qestsupplemental}
\end{align}
where we have expressed the quantum estimators in terms of Pauli matrices to better visualise their structure. Introducing now Eq.\hspace{0.2em}(\ref{labeleffstate}) and Eq.\hspace{0.2em}(\ref{qestsupplemental}) in the single-shot bound derived in this work we find that
\begin{align}
\bar{\epsilon}_{\mathrm{mse}} &\geqslant \sum_{i=1}^2 w_i \left[\int d\theta_1 d\theta_2 p(\theta_1, \theta_2)\theta_i^2 - \mathrm{Tr}\left(\rho S_i^2\right) \right]
\nonumber \\
&= \frac{1}{2} \sum_{i=1}^2  \left[ \frac{4}{\pi^2}\int_{-\pi/4}^{\pi/4} d\theta_1  \int_{-\pi/4}^{\pi/4} d\theta_2 \hspace{0.15em}\theta_i^2 - \mathrm{Tr}\left(\rho S_i^2\right) \right]
\nonumber \\
&= \frac{\pi^2}{48} - \frac{2\left(4-\pi\right)^2\left[2-\left(4-\pi^2\right)\gamma^2 + 2\gamma^4 \right]}{\pi^4 \left(1+\gamma^2\right)^2},
\label{qnetworksingleshotsupplemental}
\end{align}
which is the error for $\mu = 1$ examined in the main text.

On the other hand, if we choose $\gamma = 1$, which is one of the values that gives the minimum single-shot error, then the quantum estimators in Eq.\hspace{0.2em}(\ref{qestsupplemental}) become
\begin{eqnarray}
S_1 = \frac{\left(4-\pi\right)}{\pi\sqrt{2}}\sigma_y\otimes\mathbb{I},~~S_2 = \frac{\left(4-\pi\right)}{\pi\sqrt{2}}\mathbb{I}\otimes\sigma_y.
\label{qestloc}
\end{eqnarray}
It is thus clear that the tensor products of the eigenvectors of $\sigma_y$ form a common set of eigenvectors. This is how we arrive at the strategy in the main text given by the projectors $\ket{s_+, s_+}$, $\ket{s_-, s_-}$, $\ket{s_+, s_-}$, $\ket{s_-, s_+}$, with $\ket{s_\pm} = (\ket{0}\pm i\ket{1})/\sqrt{2}$.

\section{Calculations for phase imaging}\label{imaging1}

In this work we have considered a quantum imaging protocol based on the unitary \cite{chiara2003, humphreys2013, knott2016local}
\begin{align}
U(\boldsymbol{\theta}) &= \mathrm{exp}\left( - i \sum_{k=1}^d a_k^\dagger a_k \theta_k \right) = \mathrm{exp}\left( - i \sum_{k=1}^d N_k \theta_k \right) 
\nonumber \\
&\equiv \mathrm{exp}\left( - i \boldsymbol{N}\cdot\boldsymbol{\theta} \right),
\label{unitarysupplemental}
\end{align}
and we have assumed a flat prior with hypervolume $(2\pi/\bar{n})^d$ and centre $\bar{\theta} = (0, 0, \dots)$. For the global strategy 
\begin{equation}
\ket{\psi_0} = \frac{1}{\sqrt{d + \alpha^2}}\left(\alpha \ket{\bar{n}_0} + \sum_{k=1}^d \ket{\bar{n}_k}\right),
\label{globalschemesupplemental}
\end{equation}
where $\ket{\bar{n}_j} \equiv \ket{0}_0 \otimes \cdots \otimes \ket{0}_{j-1} \otimes \ket{\bar{n}}_j \otimes \ket{0}_{j+1}\otimes \cdots \ket{0}_d = \ket{0\dots 0~\bar{n}~0 \dots 0}$, we have that
\begin{align}
\rho &= \left(\frac{\bar{n}}{2\pi}\right)^d \int_{-\pi/\bar{n}}^{\pi/\bar{n}} d\theta_1 \cdots \int_{-\pi/\bar{n}}^{\pi/\bar{n}} d\theta_d \hspace{0.15em}\mathrm{e}^{-i \boldsymbol{N}\cdot\boldsymbol{\theta}} \ketbra{\psi_0}\mathrm{e}^{i \boldsymbol{N}\cdot\boldsymbol{\theta}}
\nonumber \\
&= \frac{1}{d+\alpha^2}\left(\alpha^2\ketbra{\bar{n}_0} + \sum_{k=1}^d \ketbra{\bar{n}_k}\right),
\label{imagingrho}
\end{align}
and that
\begin{align}
\bar{\rho}_k &= \left(\frac{\bar{n}}{2\pi}\right)^d \int_{-\pi/\bar{n}}^{\pi/\bar{n}} d\theta_1 \cdots \int_{-\pi/\bar{n}}^{\pi/\bar{n}} d\theta_d \hspace{0.15em} \mathrm{e}^{-i \boldsymbol{N}\cdot\boldsymbol{\theta}} \ketbra{\psi_0}\mathrm{e}^{i \boldsymbol{N}\cdot\boldsymbol{\theta}}\theta_k
\nonumber \\
&= \frac{-i\alpha}{\bar{n}\left(d + \alpha^2 \right)} \left(\ketbra{\bar{n}_k}{\bar{n}_0} - \ketbra{\bar{n}_0}{\bar{n}_k}\right),
\label{imagingrhomean}
\end{align}
having used the fact that 
\begin{align}
&\int_{-\pi/\bar{n}}^{\pi/\bar{n}} d\theta_j = \frac{2\pi}{\bar{n}},
\nonumber \\
&\int_{-\pi/\bar{n}}^{\pi/\bar{n}} d\theta_j\hspace{0.15em} \theta_j = \int_{-\pi/\bar{n}}^{\pi/\bar{n}} d\theta_j \hspace{0.15em} \mathrm{e}^{\pm i \bar{n}\theta_j}  = 0,
\nonumber \\
&\int_{-\pi/\bar{n}}^{\pi/\bar{n}} d\theta_j \hspace{0.15em} \mathrm{e}^{\pm i \bar{n}\theta_j} \theta_j = \pm \frac{2i \pi}{\bar{n}^2}.
\end{align}

Next we need to solve $S_k \rho + \rho S_k = 2\bar{\rho_k}$. If we decompose $\rho$ in Eq.\hspace{0.2em}(\ref{labeleffstate}) as $\rho = \sum_i p_i \ketbra{\phi_i}$ and insert it in $S_i\rho + \rho S_i = 2\bar{\rho}_i$, then we can rewrite $S_i$ as \cite{jesus2018}
\begin{equation}
S_k = 2 \sum_{j,l} \frac{\bra{\phi_j}\bar{\rho_k}\ket{\phi_l}}{p_j + p_l}\ketbra{\phi_j}{\phi_l}.
\label{multiqestdiag}
\end{equation}
By observing that $\rho$ in Eq.\hspace{0.2em}(\ref{imagingrho}) is already diagonal, Eq.\hspace{0.2em}(\ref{multiqestdiag}) simply becomes
\begin{equation}
S_k =  \frac{- 2i\alpha}{\bar{n}\left(1 + \alpha^2\right)} \left(\ketbra{\bar{n}_k}{\bar{n}_0} - \ketbra{\bar{n}_0}{\bar{n}_k}\right)
\label{qestimgsupplemental}
\end{equation} 
after using Eq.\hspace{0.2em}(\ref{imagingrhomean}). 

The results for $\rho$ and $S_k$ can be now inserted in our single-shot bound, recovering in this way the bound in the main text, that is,
\begin{align}
\bar{\epsilon}_{\mathrm{mse}} &\geqslant  \sum_{k=1}^d w_k \left[\int d\boldsymbol{\theta} p(\boldsymbol{\theta})\theta_k^2 - \mathrm{Tr}\left(\rho S_k^2\right) \right]
\nonumber \\
&= \frac{1}{d} \sum_{k=1}^d  \left[\left(\frac{\bar{n}}{2\pi}\right)^d \int_{-\pi/\bar{n}}^{\pi/\bar{n}} d\theta_1 \cdots \int_{-\pi/\bar{n}}^{\pi/\bar{n}} d\theta_d \hspace{0.15em}\theta_i^2 - \mathrm{Tr}\left(\rho S_k^2\right) \right]
\nonumber \\
&= \frac{1}{\bar{n}^2} \left[\frac{\pi^2}{3} - \frac{4\alpha^2}{\left(1 + \alpha^2\right)\left(d + \alpha^2\right)} \right].
\end{align}

Regarding the local strategy $\rho_0 = \rho_0^{\mathrm{ref}}\otimes \rho_0^{(1)}\otimes \cdots \otimes \rho_0^{(d)}$, with $\rho_0^{(k)} = |\phi_0^{(k)}\rangle \langle \phi_0^{(k)}|$ and
\begin{equation}
|\phi_0^{(k)}\rangle = \left[\sqrt{1- \frac{\bar{n}}{N(d+1)}}\ket{0}+\sqrt{\frac{\bar{n}}{N(d+1)}}\ket{N}\right] \equiv \ket{\phi_0},
\label{localstrategysupplemental}
\end{equation}
the single-shot bound can be written as
\begin{align}
\bar{\epsilon}_{\mathrm{mse}} &\geqslant   \frac{1}{d} \sum_{k=1}^d  \left[\left(\frac{\bar{n}}{2\pi}\right)^d \int_{-\pi/\bar{n}}^{\pi/\bar{n}} d\theta_1 \cdots \int_{-\pi/\bar{n}}^{\pi/\bar{n}} d\theta_d \hspace{0.15em}\theta_i^2 - \mathrm{Tr}\left(\rho S_k^2\right) \right]
\nonumber \\
&\equiv \frac{1}{\bar{n}^2} \left[\frac{\pi^2}{3} - f\left(N, \bar{n}, d\right) \right].
\end{align}
Since the prior that we are using is separable, in this case we have that 
\begin{align}
\rho &= \rho_0^{\mathrm{ref}}\otimes \rho^{(1)}\otimes  \cdots \otimes \rho^{(d)}, 
\nonumber \\
\bar{\rho}_k &= \rho_0^{\mathrm{ref}}\otimes \rho^{(1)}\otimes  \cdots \otimes \bar{\rho}^{(k)} \otimes  \cdots \otimes \rho^{(d)}, 
\nonumber \\
S_k &= \mathbb{I}_{\mathrm{ref}}\otimes\mathbb{I}\otimes \cdots \otimes S^{(k)} \otimes\cdots \otimes \mathbb{I},
\end{align}
where the single-mode operators $\rho^{(k)}$, $\bar{\rho}^{(k)}$ and $S^{(k)}$ are identical for all the modes, and the calculation of the optimal single-shot uncertainty for the local estimation of several phases is effectively reduced to the single-parameter calculation 
\begin{equation}
f\left(N, \bar{n}, d \right) = \frac{\bar{n}^2}{d} \sum_{k=1}^d \mathrm{Tr}\left(\rho S_k^2\right) = \bar{n}^2 \hspace{0.15em} \mathrm{Tr}\left(\varrho S^2\right),
\label{localfunction}
\end{equation}
where $\rho^{(k)} \equiv \varrho$ and $S^{(k)}\equiv S$. Performing calculations analogous to those in previous examples (and similar to those in \cite{jesus2018} for single-parameter NOON states), we find
\begin{align}
\varrho &= \frac{1}{\pi N\left(1+\nu^2\right)}
\begin{pmatrix}
\nu^2 \pi N & \bar{n}\nu\hspace{0.15em}\mathrm{sin}\left(N\pi/\bar{n}\right) \\
\bar{n}\nu\hspace{0.15em}\mathrm{sin}\left(N\pi/\bar{n}\right) & \pi N
\end{pmatrix},
\nonumber \\
S &= \frac{2\nu \left[N \pi \hspace{0.15em}\mathrm{cos}\left(N\pi/\bar{n}\right) - \bar{n}\hspace{0.15em}\mathrm{sin}\left(N\pi/\bar{n}\right) \right] }{\pi N^2 \left(1+\nu^2\right)} 
\begin{pmatrix}
0 & -i \\
i & 0
\end{pmatrix}
\end{align}
for the state in Eq.\hspace{0.2em}(\ref{localstrategysupplemental}), where $\nu = \sqrt{N(d+1)/\bar{n}-1}$. Therefore, Eq.\hspace{0.2em}(\ref{localfunction}) implies that
\begin{equation}
f\left(N, \bar{n}, d \right) = \frac{4\bar{n}^3\left[\left(1 + d\right) N - \bar{n} \right]a\left(N, \bar{n}\right)^2}{\pi^2 N^6 \left(1 + d\right)^2},
\end{equation}
where $a\left(N, \bar{n} \right) = N \pi \hspace{0.15em}\mathrm{cos}\left(N\pi/\bar{n}\right) - \bar{n}\hspace{0.15em}\mathrm{sin}\left(N\pi/\bar{n}\right)$, and as we mentioned in the main discussion, $f(N = \bar{n}, \bar{n}, d) = 4d/(1+d)^2$ and $f(N \gg 1, \bar{n}, d) \approx 0$.

\section{Performance of a phase imaging measurement versus our quantum bound}\label{imaging2}

Consider the global phase imaging scheme in Eq.\hspace{0.2em}(\ref{unitarysupplemental}) and Eq.\hspace{0.2em}(\ref{globalschemesupplemental}), and let us calculate the single-shot bound for $d = 2$, $\bar{n} = 2$, $\mathcal{W} = \mathbb{I}/2$, the same two-parameter prior probability that we employed in the qubit case, and $\alpha = 1$, where the latter is the balanced version of Eq.\hspace{0.2em}(\ref{globalschemesupplemental}) \cite{knott2016local}. With this configuration we find that 
\begin{align}
\rho &= \frac{1}{3}\left[\mathbb{I} + \frac{2(\lambda_1 + \lambda_4)}{\pi} + \frac{4\lambda_6}{\pi^2} \right],
\nonumber \\
\bar{\rho}_1 &= \frac{1}{3\pi}\left(\frac{2\lambda_7}{\pi}-\lambda_2\right),
\nonumber \\
\bar{\rho}_2 &= -\frac{1}{3\pi}\left(\frac{2\lambda_7}{\pi}+\lambda_5\right),
\end{align}
where $\lambda_i$ are Gell-Mann matrices \cite{gellmann1962}. Furthermore, introducing these results in $S_i\rho + \rho S_i = 2\bar{\rho}_i$ we find that the quantum estimators are
\begin{align}
S_1 &=  \frac{1}{\pi}\left[ \frac{\lambda_5-(1+\pi^2)\lambda_2}{2+\pi^2}  +\frac{\lambda_7}{\pi} \right],
\nonumber \\
S_2 &= \frac{1}{\pi}\left[ \frac{\lambda_2 -(1+\pi^2) \lambda_5}{2+\pi^2}  -\frac{\lambda_7}{\pi} \right],
\end{align}
and the single-shot error is bounded as
\begin{equation}
\bar{\epsilon}_{\mathrm{mse}} \geqslant \frac{\pi^2}{48} - \frac{2\left(4 + 3\pi^2 + \pi^4\right)}{3\pi^4\left(2 + \pi^2\right)} \approx 0.130.
\label{photonbound}
\end{equation}

While we cannot extract an optimal measurement from $S_1$ and $S_2$ because $[S_1, S_2] \neq 0$, a numerical search by trial and error has revealed an approximate set of projectors with a precision almost as good as that given in Eq.\hspace{0.2em}(\ref{photonbound}). In particular, if we use 
\begin{eqnarray}
\bra{\varphi_a}&=& (0.485 + 0.131 i, 0.441 - 0.070 i, -0.223 + 0.706 i),
\nonumber \\
\bra{\varphi_b} &=& (0.688, - 0.208 - 0.432 i, -0.270 - 0.472 i),\nonumber \\
\bra{\varphi_c}&=& (0.509 + 0.118 i, -0.284 + 0.700 i, 0.396 ),
\label{numscheme}
\end{eqnarray}
with components labelled as $\ket{2,0,0}$, $\ket{0,2,0}$, $\ket{0,0,2}$, and we employ the multi-parameter numerical algorithm in \cite{jesus2019thesis}, then we find that, for $\mu = 1$, $\bar{\epsilon}_{\mathrm{mse}} \approx 0.142$. 

Remarkably, this result suggests that, at least in this particular case, our multi-parameter bound is providing most of the information associated with the quantum part of the problem. Whether this is also the case in other configurations should be object of some future and highly interesting work.

\end{document}